**Imaging of weak phase objects with a Zernike phase plate**

C. J. Edgcombe
Dept of Physics, University of Cambridge, JJ Thomson Ave, Cambridge CB3 0HE, UK
Email: cje1@cam.ac.uk; phone +44 1223 37473

**Abstract**
Analysis of the imaging of some simple distributions of object phase by a phase plate of Zernike type shows that sharp transitions in the object phase are well transmitted. The low-frequency components of the complete object function are attenuated by the plate. The behaviour can be characterised by a cut-on parameter defined as the product of the cut-on frequency of the plate and a characteristic dimension of the object. When this parameter exceeds a value of the order of unity, a sharp boundary in the object is imaged by a Zernike plate as a dark lining inside the boundary with a white outline or halo outside the boundary, in agreement with reported observations. The maximum diameter of objects that can be imaged accurately is inversely proportional to the diameter of the hole for beam transmission in the phase plate.

**Keywords**
Phase plate; Zernike plate; weak phase object; Fourier optics.

## 1. Introduction

Phase plates are currently of interest in transmission electron microscopy (TEM) for improving the imaging of weak phase objects. When inserted at a suitable position in the column, a phase plate can change the phase of a specific range of spatial frequencies and provide maximum image contrast for these frequencies on passing through focus, thereby eliminating their contrast reversal and simplifying the interpretation of images. Use of a phase plate is also expected to reduce the electron dose needed for imaging and thus reduce specimen damage.

One frequently used type of plate is rotationally invariant and is described as having Zernike geometry. A simple form of Zernike plate consists of a thin sheet of material of known mean inner potential and provided with a central hole of radius $r_1$ somewhat greater than that of the unscattered electron beam. The thickness of the plate is chosen to change the phase of the scattered electrons by the desired amount relative to the unscattered beam, as described by Danev and Nagayama [1]. This type of plate has an outer boundary at the maximum aperture that is convenient but has no further structure. Other possible ways have been demonstrated for producing a change of phase that is independent of rotation around the axis. One method uses electrodes arranged so that direct and scattered electrons see different potential distributions as they pass through the structure. The electrical lengths for the two paths differ, providing a phase difference which can be varied in operation, as described for example by Schultheiss et al [2]. Another method being investigated uses a thin ring carrying magnetic flux which produces a phase difference by the Aharonov-Bohm effect [3]. Such a ring can either advance or retard the phase of scattered electrons, depending on the direction of circulation of flux which can be reversed by turning over the ring in its holder. In principle, a Zernike plate provides the desired phase difference at radii down to some minimum value $r_2$. For the simple sheet, $r_2$ coincides with $r_1$ but for other types of plate the two radii differ. Electrons passing the plate at radii between $r_1$ and $r_2$ may be intercepted, and those at radii less than $r_1$ will not be changed in phase.

Phase or amplitude contrast transfer functions show how the transmission varies with frequency, but do not reveal the result of imaging an object that superimposes many spatial frequencies. By considering the imaging of strong phase objects, Beleggia has shown [4] that the optimum phase change for the plate is a function of object phase shift. Further detailed simulations [5], [6] have modelled strong phase objects and the transfer function of the objective lens.   The purpose of the present paper, in contrast, is to obtain analytic results for the imaging of weak phase objects by the phase plate alone, omitting any absorption or defects of lenses.  Although the objects to be used are much simpler than real biological specimens, the analytical descriptions allow the change of image with object size to be demonstrated clearly.  The results not only confirm that objects that are sufficiently small can be imaged accurately, but also show that the 'white halo' and other artefacts that appear with larger objects are due to the high-pass filtering action of the phase plate, independent of any lens effects.  The cut-on frequency of a Zernike plate and a characteristic dimension of the object can be combined to form a dimensionless 'cut-on parameter' whose value characterises the behaviour produced by this combination.

The process of imaging by successive Fourier transforms, as described in Sec 8.6.3 of Born & Wolf [7] and Chapter 5 of Goodman [8], is convenient for analysing the effect of a phase plate.  This process defines the imaging behaviour of a coherently-illuminated perfect lens, ignoring quadratic functions of spatial coordinates and magnification.  According to Sec. 5.3.2 of [8], this neglect of quadratic factors is acceptable provided the object is sufficiently small in comparison with the lens aperture, as is likely to be satisfied in electron microscopy.

The results reported here are expressed in terms of various special functions, as defined by Abramowitz and Stegun [9] and the NIST Digital Library of Mathematical Functions [10].  This analysis omits effects both of charging and of loss in the object or phase plate, and applies only to weak phase objects.

## 2. General phase variation

At the object, a scattering centre acts as a source of a spherical wave.  An element of this wave initially diverging at an angle $\theta$ to the axis arrives at the back focal plane (BFP) at a radius $r_s(\theta)$. We adopt the usual approximation that

$$r_s \approx \theta f$$

where f is the focal length of the objective lens. To simplify later analysis, the incident and scattered wave vectors $\mathbf{k_0}$ and $\mathbf{k}(\theta)$ are defined here to have magnitude $2\pi/\lambda$, where $\lambda$ is the electron wavelength, and their difference, $\mathbf{q}(\theta) = \mathbf{k}(\theta) - \mathbf{k_0}$, is a spatial frequency in angular measure.  From the geometry of scattering for $\theta \ll 1$, $q \approx k\theta = 2\pi\theta/\lambda$ so $q \approx 2\pi r_s/\lambda f$. Since a Zernike plate produces a phase change only for $r > r_2$, it does so only for q values greater than $2\pi r_2/\lambda f$. This threshold value of q is is denoted here by $q_0$ and is $2\pi$ times the corresponding quantity defined in [1], [5], [6] as the 'cut-on' frequency.  It will be shown below that wave components with $q < q_0$ contribute little to the contrast, and so the plate acts as a high-pass filter.

When transforms of this sort are calculated without the phase plate, the integrals have ranges of 0 to ∞ or −∞ to ∞, and suitable evaluations can be found without difficulty.  However, to model a phase plate that provides a step change of phase at the cut-on frequency, it is necessary to evaluate integrals with the cut-on frequency as one of the limits.  Very few suitable integrals are then available, and this restricts the types of object for which analytic solutions can be obtained.  A semi-analytic solution is presented here for an object that produces uniform phase change over a cylindrical radius *b*.  It is possible to find analytic solutions somewhat more easily for equivalent systems in 1D Cartesian coordinates, and two such solutions are presented here for comparison.  However, the Cartesian analysis implies that not only the object but also the phase plate is of strip form and hence these solutions do not represent the behaviour of strip objects with a rotationally invariant Zernike plate.  The spatial frequency spectra of all these objects contain components down to zero frequency and so are useful for modelling extended objects.

The mean potential in an object is more negative than the vacuum potential, so an electron travelling through matter moves with slightly greater average momentum than one of the same energy in the surrounding space. Hence, within the object, the electron wavelength is smaller, k is greater and the phase change kz is increased, relative to the same distance of transit z outside the object (Sec. 4.2 of [11]). The phase changes induced both by a typical phase object relative to the direct beam, and by a Zernike plate on the scattered wave components relative to the direct beam, are thus both positive.

Consider a wave described by
$$\psi = \exp i[kz - \omega t + A\phi]$$
where A is the amplitude of phase shift and $\phi$ is a real function of $x$ or of $r$. Different functions $\phi$ will be specified for different objects, with a maximum magnitude of 1. We make the weak-phase-object approximation by assuming that $A \ll 1$. The exponential can then be expanded to first order in A as
$$\psi \approx (1 + iA\phi) \exp i(kz - \omega t)$$
Here we consider only far-field transforms, in which the factor $\exp i(kz - \omega t)$ is approximately constant and will be omitted. The remaining complex amplitude of the total wave
$$f = 1 + i A \phi \qquad (1)$$
is a function of spatial coordinates. At any given point in the object exit plane it can be represented on a phasor diagram as in Fig. 1.

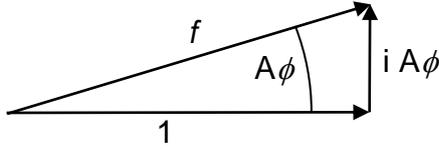

Figure 1. Wave incident on the object (shown as unit amplitude), wave $f$ at exit from the object and scattered wave iA$\phi$.

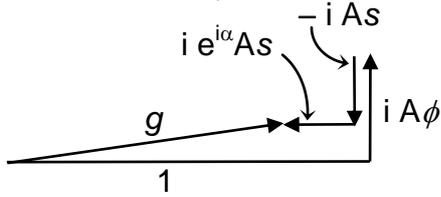

Figure 2. Wave $f$ as in Fig. 1, wave $g$ after passing through the phase plate and wave components (A$s$) introduced by the phase plate (drawn here for α = π/2).

The effect of the phase plate will be found by defining a phase function $\phi$ for the object, finding the (spatial) frequency spectrum of the whole wave, applying the phase change $\alpha$ from the plate as a function of frequency and then transforming again to find the modified wave $g$ in coordinate space. The analyses below express the results for $g$ in the form
$$g = 1 + i A \phi - i\left(1 - e^{i\alpha}\right) A s \qquad (2)$$
where $s$ is a function of a transverse coordinate and of $q_0$, the cut-on frequency of the plate. If $s$ is equal to $\phi$, then (2) gives the same result for $g$ as if the vector iA$\phi$ were rotated by $\alpha$. Just as the phase change of the original wave by the object is said to add the vector of magnitude A$\phi$, so the phase plate adds further components (A$s$) both opposing that vector and in the direction ($\alpha + \pi/2$). The resulting intensity is, from (2),
$$|g|^2 = 1 - 2 A s \sin \alpha + O(A^2)$$
Thus if the phase plate advances the scattered component (iA$\phi$) in Fig. 1 and so rotates it counter-clockwise, the total wave $g$ then becomes smaller in amplitude than the initial wave (Fig. 2), the intensity is reduced and the image becomes darker. A general rotation by $\alpha$ is analysed here, with phase advance (as provided by a thin Zernike plate) shown by a positive value of $\alpha$.

The relative distribution of intensity over the image is proportional to sin $\alpha$ which is constant over the image, and to $s$ which varies over the image. When the object is sufficiently large (to be quantified later), |s| is small over most of the image and the variation of intensity remains proportional to $A^2$. When the object is sufficiently small, the magnitude |s| increases to the same

order as $\phi$ and the intensity contains a component proportional to A. It will be shown that the distribution of $s$ in the transverse plane may differ substantially from that of $\phi$.

### 3. 1D Analysis for Cartesian geometry

The transverse coordinates of object and image are denoted here by $x$ and $x'$. In accordance with Abbe theory [7], we use for the second Fourier transform a further forward transform. To maintain unit amplitude of the wave, it is then necessary to include a factor $1/\sqrt{2\pi}$ in each forward transform (and in the inverse transform if used). When a phase plate is absent, the result of this process is in general to reproduce the object wave $\phi(x)$ at the image as $\phi(-x')$, thus demonstrating the inversion produced by a perfect lens. In the cases considered here, the functions $\phi$ are even in $x$, so the images are also even in $x'$.

The Fourier transform of a general object wave in the form of (1), giving the spectrum as a function of $q$, the spatial frequency in the x-direction, is

$$F(q) = (1/\sqrt{2\pi}) \int_{-\infty}^{\infty} (1 + iA\phi(x)) \exp{-iqx} \, dx$$
$$= \sqrt{2\pi}\, \delta(q) + i\, A\, F'(q)$$

where 
$$F'(q) = (1/\sqrt{2\pi}) \int_{-\infty}^{\infty} \phi(x) \exp{-iqx} \, dx \qquad (3)$$

and 
$$\phi(x) = (1/\sqrt{2\pi}) \int_{-\infty}^{\infty} F'(q) \exp{iqx} \, dq$$

The transmission function for the phase plate is denoted here by $T(q, q_0)$. The Zernike plate is assumed to leave unchanged those spatial frequencies with magnitude less than the transition frequency $q_0$, but to advance by $\alpha$ the phase of components with greater frequencies. Its transmission function is thus

$$T(q, q_0) = \begin{cases} \exp i\alpha, & |q| > q_0 \\ 1, & |q| < q_0 \end{cases} \qquad (4)$$

In practice, the transmission of the plate is zero above some upper limit of $q$ that is much greater than $q_0$. This could be taken into account by modifying one limit of the functions $s$ defined below. The effect is small for the frequency range of interest here and in the second transforms below the upper limit of $q$ has been extended to infinity. Applied to the wave $F(q)$, $T$ modifies the spectrum to $G(q,q_0)$:

$$G(q, q_0) = F(q)\, T(q, q_0)$$

The image is then a further transform of $G$, in which the new transverse coordinate is denoted by $x'$:

$$g(x', q_0) = (1/\sqrt{2\pi}) \int_{-\infty}^{\infty} F(q)\, T(q, q_0) \exp(-iqx') \, dq$$
$$= 1 + i\,(A/\sqrt{2\pi}) \left[ e^{i\alpha} \int_{-\infty}^{-q_0} F'(q) \exp(-iqx') \, dq + \int_{-q_0}^{q_0} + e^{i\alpha} \int_{q_0}^{\infty} \right] \qquad (5)$$
$$= 1 + i\, A\, \phi(-x') - i\,(1 - e^{i\alpha}) A\, s$$

where in (5) all the integrals have the same integrand, and $s$ is defined by

$$s = \sqrt{(2/\pi)} \int_{q_0}^{\infty} \text{Re}[F'(q) \exp(-iqx')] \, dq \qquad (6)$$

The $\delta$-function at zero frequency in $F(q)$ is not altered by the phase plate and so the unscattered part of the wave reappears in $g$. In the functions to be considered, $\phi(x)$ is even in $x$ and $F'(q)$ is real.

### 3.1 Rectangular (or top-hat) distribution of phase

Consider an object phase distribution of unit magnitude over a width of 2b centred on $x$ = 0:

$$\phi(x) = \begin{cases} 1, & |x| < b \\ 0, & |x| > b \end{cases}$$

For this distribution, $F'(q)$ can be obtained from (3) and substituted into (6). The result is defined as $s_1(x')$:

$$s_1(x') = (1/\pi) \int_{q_0}^{\infty} [\sin q(x' + b) - \sin q(x' - b)]/q \; dq \quad (7)$$

The integral for $s_1$ with a non-zero lower limit can be evaluated by use of the sine-integral function Si described in [9] and [10]. The result depends on the relative magnitudes of $x'$ and b, as shown in Appendix A:

$$s_1(x') = \begin{cases} \sigma, & |x'| > b \\ 1 + \sigma, & |x'| < b \end{cases}$$

where $\quad \sigma = (1/\pi)[\text{Si}(q_0(x' - b)) - \text{Si}(q_0(x' + b))]$

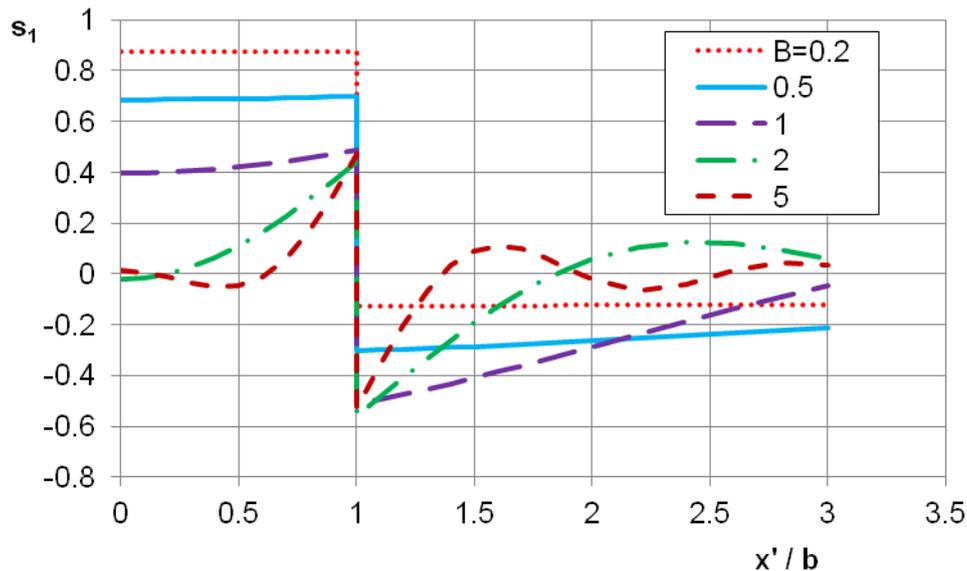

Figure 3. Variation of image intensity produced by a π/2 phase plate of cut-on angular frequency $q_0$ in Cartesian geometry. Here the object phase is assumed to be uniform from x = –b to +b, $x'$ is the transverse coordinate at the image, the horizontal axis is $X = x'/b$ and the parameter is B = $q_0 b$.

The function $s_1$ is plotted in Fig. 3 as a function of $x'/b$, and in Fig. 4 as a function of $q_0 x'$. For both plots, the parameter B is $q_0 b$. Fig. 4 shows the response for a fixed $q_0$ to object functions of varying widths. The object function, if plotted, would appear as a rectangle of unit amplitude, changing to zero at $x' = b$. Positive or negative values of $s_1$ imply that the total intensity is respectively less than or greater than the background (for positive α). The value of $s_1$ on the axis ($x' = 0$) is $[1 - (2/\pi) \text{Si}(B)]$ which, as B increases from zero, initially falls smoothly and then oscillates around zero when B > 1.92.

It can be seen in each plot that the size of the discontinuous step in phase at $x' = b$ is imaged accurately, but the imaging of phase at other values of $x'$ depends strongly on B. This dependence on B is discussed later with the results for other objects.

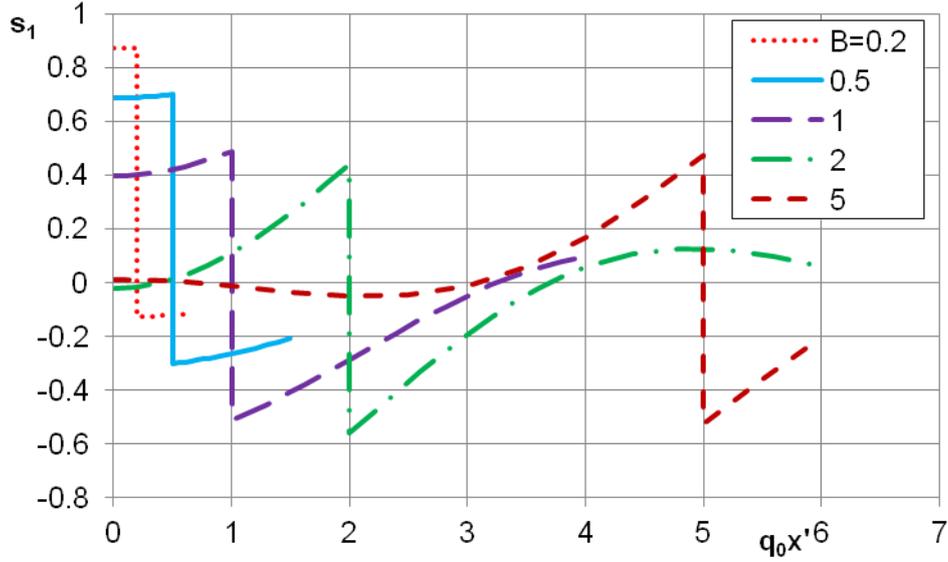

Figure 4. Variation of image intensity for the same conditions as for Fig. 3, but plotted as a function of $q_0 x'$. This axis is proportional to the transverse coordinate $x'$, for a fixed value of $q_0$.

### 3.2 Gaussian distribution of phase

Consider the distribution

$$\phi(x) = \exp(-v^2(x))$$

where $\quad v(x) = x/\sigma\sqrt{2}$

and $\sigma$ is the standard deviation of the Gaussian, assumed to be known. The transform $F'(q)$ is, from (3),

$$F'(q) = \sigma \exp(-u^2)$$

where $\quad u = \sigma q_0 /\sqrt{2}$

By using the indefinite integral given as equation (7.4.32) of [9], with $a = \sigma^2/2$, $b = \pm ix'/2$, $c = 0$, we can evaluate (6) to obtain $s_2$ for this object as

$$s_2(u, v(x')) = \phi(x') \, \text{Re}\left[\text{erfc}(u + i \, v(x'))\right]$$

A close approximation for $\text{erf}(x + iy)$ is given by eqn. (7.1.29) of [9], from which

$$s_2(u,v) \approx e^{-v^2}\left(\text{erfc}\, u - e^{-u^2}\left[(1 - \cos 2uv)/2\pi u + \frac{2}{\pi}\sum_{n=1}^{\infty} e^{-n^2/4} f_n(u,v)/(n^2 + 4u^2)\right]\right)$$

(8)

with $\quad f_n(u,v) = 2u - 2u \cosh nv \cos 2uv + n \sinh nv \sin 2uv$

At values of $u$ and $v$ greater than 1, $f_n$ can be much greater than 1, but in the expression for $s_2$, $f_n$ is reduced by the factors $\exp(-u^2)$ and $\exp(-v^2)$.

The function $s_2$ is plotted as a function of $v$ in Fig. 5 and of $q_0 x' = 2 u*v$ in Fig. 6, for $u = 0.01$, 0.1, 0.2, 0.5, 1 and 2, using values of $n$ up to 10 in (8). The Gaussian form for the phase of the object is similar to the plot for $u = 0.01$. The object distribution is reproduced with reasonable fidelity only for values of $u$ less than about 0.1. This limit occurs because for the phase plate to be effective, it must act over most of the frequency range of the object, which extends to about $|u| = \sqrt{2}$. Thus, using a plate with a given value of $q_0$, we can expect to image accurately Gaussian objects for which $\sigma$ is less than about $0.14 / q_0$.

For an object of this type, the magnitude of $s_2$ on the axis ($v = 0$) is erfc($u$) and falls steadily to zero as $u$ increases.

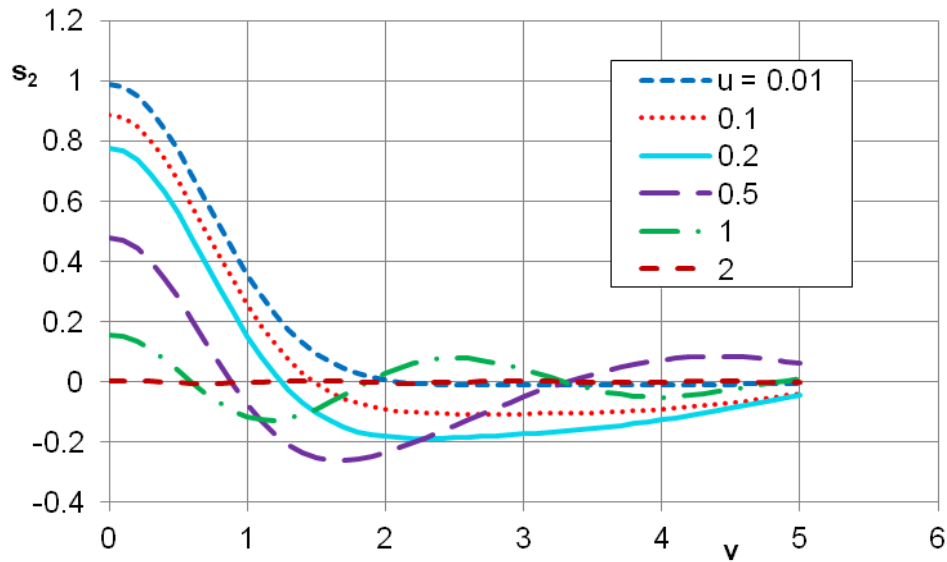

Figure 5. Variation of image intensity produced by a $\pi/2$ phase plate of transition frequency $q_0$ in Cartesian geometry. Here the object phase distribution is assumed to be Gaussian, of form $\phi = \exp -v^2$ where $v = x'/(\sigma\sqrt{2})$ and the standard deviation of the Gaussian is $\sigma$. The parameter $u$ is related to the transition frequency $q_0$ by $u = \sigma q_0 / \sqrt{2}$.

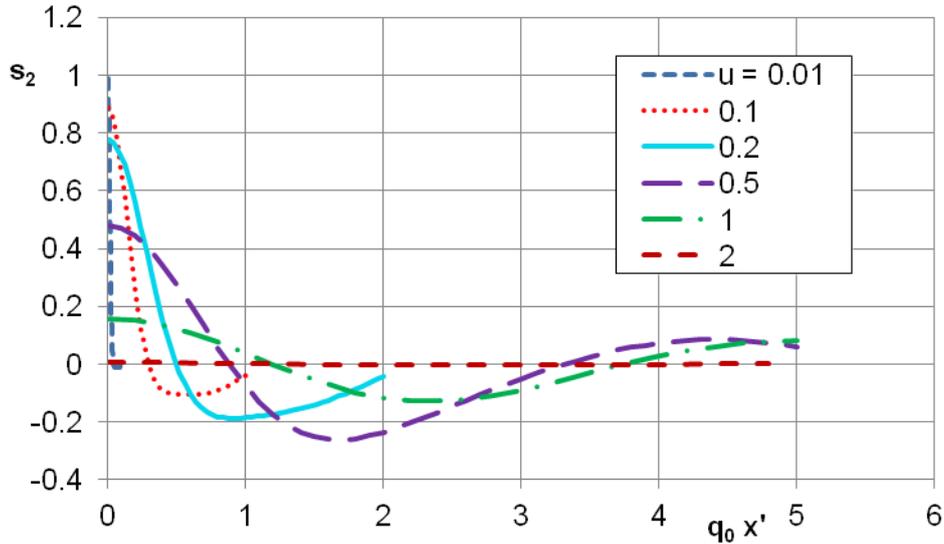

Figure 6. Variation of image intensity for the same conditions as for Fig. 5, but plotted as a function of $q_0 x' = 2 u*v$.

### 4. Rotationally invariant object

The extension of the Fourier transform to two transverse Cartesian coordinates is well known. For consistency with the relations above, a factor $(1/2\pi)$ is included in the transform:

$$F(q_x, q_y) = (1/2\pi) \int_{-\infty}^{\infty} \int_{-\infty}^{\infty} f(x,y) \exp -i(q_x x + q_y y) \, dx dy \qquad (9)$$

When this expression is converted to polar coordinates ($r$, $\theta$), the element of integration becomes ($r\,dr\,d\theta$). By defining
$$q^2 = q_x^2 + q_y^2, \quad r^2 = x^2 + y^2, \quad \tan\theta = y/x \text{ and } \tan\gamma = q_y/q_x$$
the exponent in (9) can be expressed as
$$-i(q_x x + q_y y) = -iqr\cos(\theta - \gamma)$$
For a **rotationally invariant** object, the integration over $\theta$ then yields, by eqn. (9.1.19) of [9],
$$\int_{-\pi}^{\pi} \exp[-iqr\cos(\theta - \gamma)]d\theta = 2\pi J_0(qr)$$
where $J_0()$ is a Bessel function of the first kind. On insertion into (9), the transform for a rotationally invariant function $\phi(r)$ can then be written as
$$F_0(q) = \int_0^\infty \phi(r) J_0(qr) r\, dr$$
This relation is one half of a transform pair of which the other half is
$$\phi(r) = \int_0^\infty F_0(q) J_0(qr) q\, dq$$
This pair has been described as 'the Hankel transform of order zero' and also as a Fourier-Bessel transform. (Some care is needed in using listed transforms since definitions with other weighting functions have also been described as Hankel transforms.)

Consider an object exit wave whose phase is rotationally invariant but varies with radius. The wave has the form
$$\psi = \exp i(kz - \omega t + A\phi(r))$$
As above, we assume that A is small compared to 1, expand the exponential as
$$\psi \approx (1 + iA\phi(r)) \exp i(kz - \omega t)$$
and omit the factor $\exp i(kz - \omega t)$.

### 4.1 Rectangular (top-hat) distribution

Here we consider a rectangular distribution of phase, of the same form as for Cartesian geometry but now a function of r:
$$\phi(r) = \begin{cases} 1, |r| < b \\ 0, |r| > b \end{cases}$$
The Hankel transform of $(1 + iA\phi)$ is
$$F_0(q) = \int_0^\infty J_0(qr) r\, dr + iA \int_0^b J_0(qr) r\, dr$$
$$= \delta(q)/q + i A b^2 J_1(qb)/qb$$
The second term in this result is the well-known Airy distribution.

The transmission of a phase plate with phase change $\alpha$ at radii greater than $q_0$ is specified as in (4) for the Cartesian systems (but now with q varying radially so for a conventional Zernike plate). We insert this in the spectrum at exit from the plate, write a further Hankel transform with the new radial coordinate denoted by $r'$ and insert the known integral over the full range (eqn. 11.4.42 of[9]). Then on defining
$$t = b \int_0^{q_0} J_1(qb) J_0(qr')\, dq \tag{10}$$
(where the limits are chosen to avoid numerical evaluation at infinity) and
$$s_3 = \phi(r') - t$$
the result can be written in the same form as for the Cartesian systems:

$$g(r', q_0) = 1 + i A \phi(r') - i(1 - e^{i\alpha}) A s_3$$

The integral for $t$ may be found indirectly as follows. From the definition of $t$,

$$\partial t/\partial r' = -b \int_0^{q_0} J_1(qb) J_1(qr') q \, dq$$

An indefinite integral (eqn. 11.3.29 of [9]) can now be used, leading to

$$\partial t/\partial \rho = B [\rho J_1(B) J_0(\rho B) - J_0(B) J_1(\rho B)]/(\rho^2 - 1) \qquad (11)$$

where $\quad \rho = r'/b \quad and \quad B = q_0 b$

The values of $t$ and $s_3$ on the axis can be found from (10):

$$t(r' = 0) = 1 - J_0(B)$$
$$s_3(r' = 0) = J_0(B)$$

and equation (11) can then be integrated numerically with $\rho$ increasing from zero.

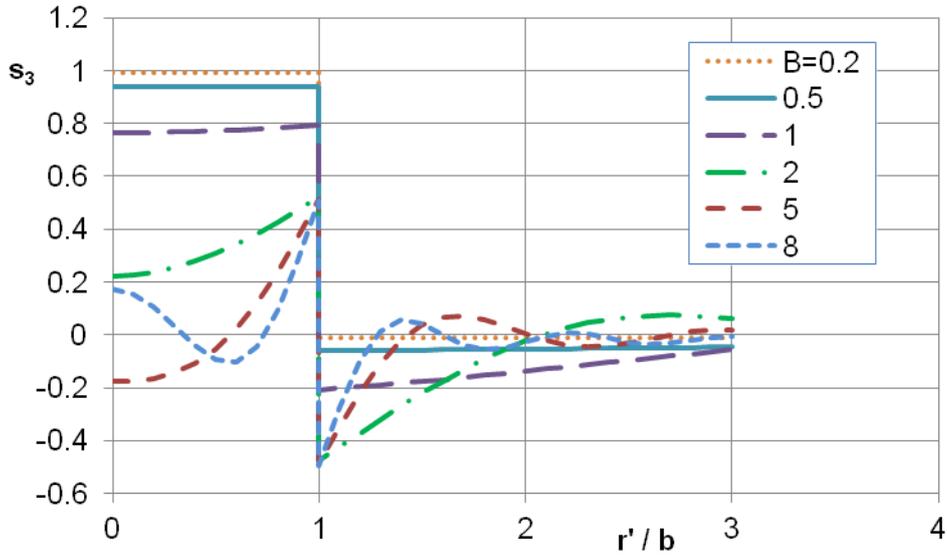

Figure 7. Variation of image intensity produced by a π/2 phase plate of cut-on angular frequency $q_0$ in a rotationally symmetric system. Here the object phase is assumed to be uniform within a radius $b$, the horizontal axis is ($r'/b$) and the parameter is B = $q_0 b$.

Fig. 7 shows $s_3$ (obtained by trapezoidal approximation with step size 0.1) as a function of $r'/b$ for B = 0.2, 0.5, 1, 2, 5 and 8, and Fig. 8 shows $s_3$ as a function of $q_0 r'$ for the same values of B. These results are similar to those of Figs. 3 and 4 for the Cartesian rectangular object function, but with the difference that for the circular object the values of $s_3$ on the axis do not drop steadily to zero as B increases but oscillate as $J_0(B)$.

Fig. 8 compares the imaging of objects of different diameters by a plate with a sharp cut-on at angular frequency $q_0$. The steps of sharp phase transition are imaged fully. Objects may be imaged accurately if B is less than about 1 (see Discussion).  As B increases, the low-frequency components of the object function are progressively lost from the image, the mean value at a transition falls to the background value ($s_3$ = 0), the overshoot outside the object's boundary increases and the contrast within the boundary correspondingly decreases and oscillates as a function of both B and $r'$.

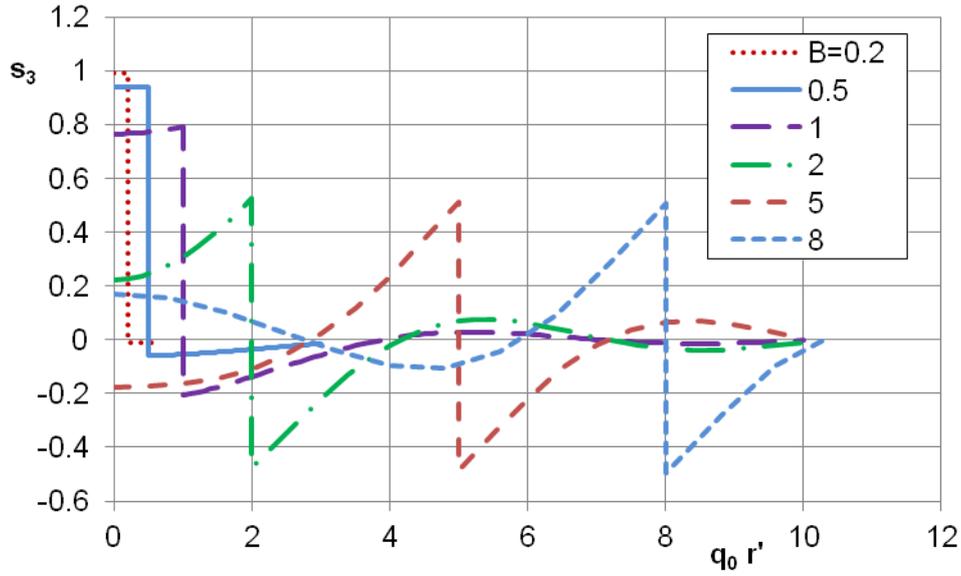

Figure 8. Variation of image intensity for the same conditions as for Fig. 7, but plotted as a function of $q_0 r'$.

## 5. Discussion

The image profiles presented here show the effect of the phase plate alone, without influence from any lens, electron absorption in the plate or charging.

In general the behaviour is similar for all three phase distributions considered. Where these objects have a sharp transition in phase at an interface or boundary, the magnitude of the transition is accurately transferred to the image since its spectrum extends to infinitely high frequencies. The response to other features of these objects depends on a dimensionless parameter B given by the product of spatial frequency and a characteristic dimension of the object. An image intensity that is everywhere proportional to the phase shift in the object is obtained only when the 'cut-on parameter' defined in this way is smaller than a value of the order of unity. As this parameter is increased to greater values, the mean value at a sharp transition settles to the background value ($s_i$ = 0), with overshoot outside the transition and a reduction in amplitude, together with oscillation with $x'$ or $r'$, inside the transition. For the Zernike plate consisting of a carbon disc, the intensity is brighter in the overshoot, thus producing the white halos reported by Danev and Nagayama (Fig 9(b) & (c) of [11]) and by Danev, Glaeser and Nagayama (Fig. 9(c) of [6]). Also as the cut-on parameter increases, the relative magnitude of the image inside the boundary decreases and oscillates, as a consequence of the 'high-pass' action of the Zernike plate.

Of the objects considered here, the response to a circular object shown in Figures 7 and 8 is most relevant to biological imaging. When the object is rotationally invariant, a continuous increase in $q_0$ for the plate causes the amplitude at the centre of the image not to fall smoothly to the mean, but to oscillate as $J_0(B)$. The observed intensity at the centre of the image may appear correspondingly brighter or darker than the mean, depending on the sign of $J_0(B)$. For the weak object considered here with uniform phase shift across its diameter, the intensity on the axis falls to the mean when B is about 2.40, that is when the cut-on periodicity is about 1.3 times the diameter. The curve in Fig. 8 for B = 8 shows that intensity fluctuations are possible as seen within the boundaries of phages in Fig. 1b of [6], but predicts more of a white halo than appears in that figure. As described in [6], the phages are strong phase objects.

From the calculations shown here, it seems that the imaging of an object may be acceptably accurate when B as defined here for the combination of object and phase plate is less than unity. This condition is in reasonable agreement with that found by Danev, Glaeser and Nagayama [12], who state that 'the cut-on periodicity should be at least twice the particle size'. The angular frequency $q_0$ used here is ($2\pi$/cut-on periodicity), so in terms of our $B = q_0 b$, the condition given in [12] is that $B \leq \pi/2$. Also, Hall, Nogales and Glaeser find [5] in detailed simulations that maximum contrast is achieved when the period at cut-on approaches three times the particle size, that is when $B \sim 1.05$. The condition obtained here for simple objects, that B should be less than about 1 for good fidelity, thus agrees well with that found experimentally and by simulation for more realistic objects.

The maximum object diameter, 2 $b_{max}$, that can be reliably imaged can be obtained easily from the corresponding value $B_{max}$ and a relation for $q_0$:
$$2b_{\max} = 2B_{max}/q_0 = \lambda\, B_{max}\, f/\pi r_2 \sim \lambda f/\pi r_2$$
This maximum diameter is thus inversely proportional to the smallest diameter in the Zernike plate at which phase correction is effective.

### 6. Conclusion

The imaging of some simple distributions of object phase by phase plates of Zernike type has been analysed. The results show how the high-pass behaviour of the phase plate modifies the images obtained. Where an object has a sharp transition in phase, for instance at a boundary, the high-frequency components of the transition are transmitted fully to the image. A cut-on parameter B can be defined as the product of a characteristic dimension of the object and the cut-on frequency of the phase plate. As the value of this parameter increases, low-frequency components of the object phase distribution are suppressed over an increasing range. The mean intensity within the image falls and the mean value of a transition falls towards the background value. To maintain the size of the change at the transition, the image intensity changes from darker than background just inside the boundary to lighter than background just outside it, so producing the white outline or halo that has been reported in [6] and [11].

For the objects considered, imaging is accurate for values of B up to a value $B_{max}$ of the order of unity. For given $B_{max}$, the maximum diameter of objects that will be imaged accurately is inversely proportional to the minimum effective radius of the plate. The diameter of the hole in a Zernike plate required for the passage of the direct beam, together with other parameters of the microscope, thus sets a upper limit for the diameter of object that will be well imaged. The value of $B_{max}$ found by the calculations presented here agrees well with those deduced from observations and simulations reported elsewhere.

### Acknowledgments

The author thanks A. Howie and R. M. Glaeser for stimulating discussions and C.H.W. Barnes for the provision of facilities within the TFM Group of the Cavendish Laboratory. The author declares no competing financial interest.

**Appendix A**   Evaluation of $s_1(x')$ for a rectangular phase distribution in a Cartesian system

Consider

$$y = (1/\pi) \int_{q_0}^{\infty} \sin aq/q \, dq$$

If a > 0, then (a$q$) is positive over the range and $y$ can be written using the standard sine integral [9], [10] as

$$y = (1/\pi)[\pi/2 - \text{Si}(aq_0)], \quad a > 0$$

If a < 0, then by defining a′ = −a we obtain

$$y = -(1/\pi)[\pi/2 - \text{Si}(-aq_0)], \quad a < 0$$

In (7), if $x' < -b$, both $(x'+b)$ and $(x'-b)$ are < 0 and so

$$s_1(x') = (1/\pi)\big[\text{Si}\big(-q_0(x' + b)\big) - \text{Si}\big(-q_0(x' - b)\big)\big]$$

If $-b < x' < +b$, then $(x' - b) < 0$ and $0 < (x' + b)$ so

$$s_1(x') = 1 + (1/\pi)\big[-\text{Si}\big(q_0(x' + b)\big) - \text{Si}\big(-q_0(x' - b)\big)\big]$$

If $b < x'$, then both coefficients are positive and

$$s_1(x') = (1/\pi)\big[-\text{Si}\big(q_0(x' + b)\big) + \text{Si}\big(q_0(x' - b)\big)\big]$$

These results can be summarized as

$$s_1(x') = \begin{cases} \sigma, & |x'| > b \\ 1 + \sigma, & |x'| < b \end{cases}$$

where $\quad \sigma = (1/\pi)\big[\text{Si}\big(q_0(x' - b)\big) - \text{Si}\big(q_0(x' + b)\big)\big]$